\documentclass[aps,prl,preprint]{revtex4}

\newcommand{\eg}{{\it e.g.}}
\newcommand{\mrm}[1]{\mbox{\rm #1}}
\newcommand{\be}{\begin{equation}}
\newcommand{\ee}{\end{equation}}
\newcommand{\br}{\begin{eqnarray}}
\newcommand{\bea}{\begin{eqnarray}}
\newcommand{\eea}{\end{eqnarray}}
\newcommand{\er}{\end{eqnarray}}
\newcommand{\ba}{\begin{array}}
\newcommand{\ea}{\end{array}}
\newcommand{\bi}{\begin{itemize}}
\newcommand{\ei}{\end{itemize}}
\newcommand{\bn}{\begin{enumerate}}
\newcommand{\en}{\end{enumerate}}
\newcommand{\bc}{\begin{center}}
\newcommand{\ec}{\end{center}}

\newcommand{\nn}{\nonumber\\}

\newcommand{\Eq}[1]{Eq.(\ref{#1})}
\newcommand{\rfn}[1]{(\ref{#1})}

\newcommand{\gsim}{\lower.7ex\hbox{$\;\stackrel{\textstyle>}{\sim}\;$}}
\newcommand{\lsim}{\lower.7ex\hbox{$\;\stackrel{\textstyle<}{\sim}\;$}}

\def\mysection#1{{\bf #1.} }

\begin{document}


\title{\bf 
Relation between the neutrino and quark mixing angles and grand unification 
}

\author{ Martti Raidal}

\affiliation{National Institute of Chemical Physics and Biophysics, 
Ravala 10, Tallinn 10143,  Estonia
}


\vspace*{1.in}

\begin{abstract}
We  argue that there exists simple relation between the quark 
and lepton mixings which supports the idea of grand unification
and probes the underlying robust bi-maximal fermion mixing structure of 
still unknown flavor physics. In this framework the quark mixing matrix is 
a parameter matrix describing the deviation of neutrino mixing from 
exactly bi-maximal, predicting 
$\theta_{sol}+\theta_{C}=\pi/4$, where $\theta_C$ is the Cabibbo angle,
$\theta_{atm}+\theta_{23}^{CKM}=\pi/4$
and $\theta_{13}^{MNS}\sim \theta_{13}^{CKM}\sim {\cal O}(\lambda^3) $, 
in a perfect agreement with experimental data. Both non-Abelian and Abelian 
flavor symmetries are needed for such a prediction to be realistic.
An example flavor model capable to explain this flavor mixing pattern, 
and to induce the measured quark and lepton masses, is outlined.  
\end{abstract}

\maketitle

\mysection{Introduction}
Despite of enormous experimental progress in neutrino~\cite{nuexp} and quark 
physics in recent years, the origin of flavor remain a mystery. In the 
standard model the Yukawa couplings are free parameters to be fixed from 
experimental data. Grand unified theories (GUTs)~\cite{GG,FM}, which are 
supported by the unification of gauge couplings~\cite{unification} in 
the minimal supersymmetric standard model (MSSM), 
predict relations between the quark and lepton Yukawa couplings at 
the unification 
scale. Although those predictions must be corrected in the minimal GUTs
if all three generations of particles are considered, 
the idea of grand unification has been widely accepted. 
In the context of GUTs, the structure of Yukawa couplings has been most 
commonly derived from the Froggatt-Nielsen mechanism~\cite{FN} of Abelian 
flavor symmetry breaking. This mechanism naturally predicts small mixing angles 
which are related to hierarchical fermion masses via 
$\theta_{ij}\sim \sqrt{m_i/m_j}$, $i<j,$ 
in a reasonable agreement with the experimental data on the quark 
mixing matrix (CKM)~\cite{C,KM,W}.

This picture has been challenged by the discovery of almost bi-maximal 
neutrino mixing. If the smallness of neutrino masses is explained with the
seesaw mechanism~\cite{seesaw}, hierarchical Yukawa couplings with small
off-diagonal elements must produce large neutrino mixing angles. Although this
is technically possible~\cite{AF,GN,S,K,BW}, it requires numerical fine tunings 
between Yukawa couplings of different generations~\cite{Ma}. 
In this context non-Abelian
flavor symmetries, continuous or discrete, 
can be considered better candidates for explaining the
systematics in the neutrino mixing matrix (MNS)~\cite{MNS}. 
However, even in non-Abelian flavor models some numerical coefficients must be 
fixed by hand in order to simultaneously satisfy~\cite{str} 
the exactly maximal atmospheric 
neutrino mixing, $\sin^2 2\theta_{atm}=1.00\pm 0.05$, 
large but not maximal solar neutrino mixing,  
$\tan^2\theta_{sol}=0.41\pm 0.05$, 
vanishing $\sin^2 2\theta_{13}^{MNS}=0\pm 0.065$,
and small Cabibbo angle $\theta_C$~\cite{C} 
(or the Wolfenstein parameter $\lambda$~\cite{W}), 
$\lambda=\sin\theta_{C}=0.22$.
Although the deviation of the neutrino mixing matrix from bi-maximal
has been parametrized~\cite{R}, and the numerical correlation with 
the Cabibbo mixing has been pointed out~\cite{S},
no physics explanation relating the quark and lepton 
mixings has been given so far.

In this Letter we show that there actually exists 
a simple relation between  the quark and lepton mixings 
which provides a new experimental evidence for grand unification.
We argue that at  fundamental level the underlying non-Abelian 
flavor physics is robust and admits 
only vanishing or maximal mixing angles. Indeed, with the SO(3) or 
SU(2) flavor symmetry,
and with the simplest superpotentials for flavons, this has been shown 
to be the case~\cite{BHKR}. 
Because of GUT constraints for the fermion mixing
matrices,  the quark and lepton flavor mixings are related, predicting
\bea
\label{prediction}
&& \theta_{C}+\theta_{sol}=\frac{\pi}{4},\qquad 
\theta_{23}^{CKM}+\theta_{atm}=\frac{\pi}{4}, \\
&& \theta_{13}^{MNS}\sim \theta_{13}^{CKM}\sim {\cal O}(\lambda^3), \nonumber
\eea
in a good agreement with the experimental data (see \rfn{expprediction}). 
Those predictions test experimentally  the idea of grand unification,
and, additionally, probe the fundamental properties of still unknown 
flavor physics. They allow to rule out  the proposed idea in 
forthcoming neutrino oscillation experiments (for example if $\theta_{13}^{MNS}$ 
close to the present bound will be measured).

The resulting picture is simple and predictive. 
In the Wolfenstein parametrization~\cite{W}, there is just one 
nontrivial parameter $\lambda$ characterizing both the deviation of 
the CKM matrix from diagonal matrix, and the deviation of the
neutrino mixing matrix from exactly bi-maximal. 
The non-Abelian flavor symmetry implies singular  $2\times 2$
sub-structures for the Yukawa matrices and, consequently,  a prediction
of hierarchical fermion masses.
Realistic masses for all the  fermions
should come from the additional Froggatt-Nielsen type mechanism of 
U(1) flavor symmetry breaking. Since the breaking of the non-Abelian flavor 
symmetry which generates mixing, and the Abelian  flavor symmetry which 
generates light
fermion masses are not related, it is possible to predict \rfn{prediction}
and  to generate the realistic fermion masses at the same time~\cite{BHKR}. 
Although in this picture the Cabibbo angle is a parameter measuring an 
additional rotation,
it is intriguing to argue that it is related to the breaking of the Abelian 
flavor symmetry. Such a model building  is beyond the scope of 
this paper.

\mysection{Flavor mixing and unification}
We start with discussing how the bi-maximal fermion
mixing, and the additional rotation by  $\theta_{C}$, are consistent
with the MSSM superpotential and the GUT relations for the Yukawa couplings. 
This follows by an example how such a framework can arise from the underlying
flavor physics.

The superpotential of the MSSM with singlet (right-handed) heavy neutrinos
is given by
\bea
W= 
D^c Y_d Q H_1 + U^c Y_u Q H_2 + E^c Y_e L H_1 + N^c Y_\nu L H_2 
+\frac{1}{2} N^c M N^c,  
\label{WMSSM}
\eea
where the Yukawa matrices $Y$ are $3\times 3$ matrices which can be
diagonalized by bi-unitary transformations $Y^D=U^\dagger Y V$, where
$V,\,U$ refer to the rotation of left- and right-chiral
fields, respectively (for a symmetric matrix $Y$, $U=V^*$). 
There are two types of GUT relations between the Yukawa couplings
of \Eq{WMSSM} often considered in literature. If the MSSM fermions are
assigned into multiplets according to the SU(5) gauge group, the minimal 
unified model predicts
\bea
Y_e = Y_d^T, \;\; \quad Y_u=Y_u^T .
\label{su5rel}
\eea
However, SU(5) GUTs do not include right-chiral neutrinos.
The second constraint, so called SO(10) relation~\cite{BW},
relates the up-type Yukawa couplings as
\bea
Y_\nu = Y_u\,.
\label{so10rel}
\eea
Although the comparison of   down quark and charged lepton 
masses implies that the minimal GUT relation \rfn{su5rel} has to be 
corrected~\cite{GJ,AFM}, let us assume in the beginning that 
both the relations \rfn{su5rel}, \rfn{so10rel} hold.
After that we show how the prediction \rfn{prediction} can follow from
the SU(5) relation \rfn{su5rel} {\it alone}.
After presenting our basic results we show that  \rfn{su5rel}, \rfn{so10rel}
are actually unnecessarily restrictive for us, and the light quark masses can be
realistic without spoiling the prediction \rfn{prediction}.

Integrating out the heavy singlet neutrinos from \Eq{WMSSM}, the seesaw 
mechanism~\cite{seesaw} induces the effective operator
\bea
\frac{1}{2} \kappa LLH_2H_2,
\label{effop}
\eea
which after the electroweak symmetry breaking generates masses
for the active neutrinos as $m_\nu=\kappa v^2=Y_\nu^T M^{-1} Y_\nu v^2.$
We recall that what is observed in the quark and neutrino
experiments at low energies, the quark and neutrino mixing matrices 
$V_{CKM}$ and $V_{MNS}$,
respectively, are given by 
\bea
V_{CKM}=V^\dagger_u V_d ,\quad\quad\quad\quad
V_{MNS}=V^\dagger_e V_\nu .
\label{VCKM}
\eea                           
The right-rotations are not directly observable in the present experiments.
In the following we assume that the heavy singlet neutrino mass matrix $M$
does not introduce observable mixing effects into the light neutrino mass matrix. 
It is convenient to think of the mixing matrices $U,V$ as the sequence of 
three $2\times 2$ rotations~\cite{valle1}\footnote{Counting of physical 
phases~\cite{valle2}, which is
different for Dirac and Majorana neutrinos, does not affect our discussion 
here.}, 
\bea
V_{}, U_{}=R(\theta_{23})R(\theta_{13})R(\theta_{12}) .
\label{V}
\eea
Firstly, this allows us to simplify our discussion. Secondly, we
argue that the underlying flavor physics actually generates a 
sequence of $2\times 2$ rotations, thus \Eq{V} could correspond to the real 
situation in generating the flavor.

We argue that the underlying flavor physics admits only vanishing or
maximal mixing, and that the experimental data supports this
view on flavor. The well known result~\cite{AF} is that the SU(5) relations
\rfn{su5rel} allow the maximal atmospheric neutrino mixing
and the (almost) vanishing third generation mixing in the
$V_{CKM}$ to be consistent with \rfn{WMSSM} and \rfn{VCKM}. 
Consider the relevant $2\times 2$ rotations by $\theta_{23}.$
Choosing a basis in which $Y_u, Y_\nu$ are diagonal, and working with
the precision up to first order in  $\lambda$,
$V_{CKM}={\bf 1}$ implies  $V_{d}={\bf 1}$. 
Consequently, the maximal atmospheric
mixing should come from the maximal (2-3) mixing in   $V_{e}$, which according
to \rfn{su5rel} corresponds to the unobservable maximal right-mixing $U_d$ in
the down quark sector.

The vanishing (to first order in $\lambda$) (1-3) mixing angles in
$V_{CKM}$ and $V_{MNS}$ can be obtained trivially by setting  $\theta_{13}=0$
in all the mixing matrices involved.

If we deal with the (1-2) mixing angles in the same way as we discussed the 
atmospheric neutrino mixing, we obtain exactly bi-maximal $V_{MNS}$
and diagonal $V_{CKM}.$ However, this does not correspond to reality.
In the $V_{CKM}$ the only sizable non-zero mixing angle is the Cabibbo angle, 
while in the neutrino sector the solar mixing angle is bounded to
be non-maximal by several sigmas, $\tan^2\theta_{sol}=0.41\pm 0.05$~\cite{str}.
It is intriguing that the deviation from the exact bi-maximal mixing
in $V_{MNS}$, and the deviation from the unit matrix in $V_{CKM}$ are 
correlated: both  are in $\theta_{12}$.

To make the $V_{CKM}$ realistic, let us take the previously described 
Yukawa matrices giving bi-maximal neutrino mixing and 
introduce into $Y_u$ an additional (1-2) rotation by the Cabibbo angle, $V_C$, 
\bea
Y_u \to V_C^T Y_u V_C.
\label{VC}
\eea
This implies that  $V_{CKM}=V_{C}$
in agreement with the experiment. However, because of the SO(10) GUT 
relation \rfn{so10rel}, the same rotation by $V_C$ takes also place 
in $Y_\nu$.  This, according to \Eq{VCKM}, rotates  $V_{MNS}$ into
{\it an opposite} direction and decreases the solar mixing angle by the Cabibbo
angle, $\theta_{sol}=\pi/4 - \theta_C.$ Thus the relation between the
quark and neutrino mixing comes from the GUT relation \rfn{so10rel}.
Let us see what experimental data tells about this relation. While
$\theta_C^{exp}=12.7^\circ$ with small errors, $\tan^2\theta_{sol}=0.41\pm 0.05$
implies $\theta_{sol}^{exp}=32.6^\circ\pm 1.6^\circ.$ Thus, 
\bea
\theta_{sol}^{exp}+\theta_C^{exp}= 45.3^\circ\pm 1.6^\circ
\qquad (1\sigma),
\label{expprediction}
\eea
in a perfect agreement with the prediction.
We recall that, because of tiny first generation
quark Yukawa couplings,  $\theta_C$ practically does not run
when evaluating from $M_{GUT}$ to low energies. For normally
hierarchical neutrinos predicted by GUTs, 
$m_{\nu_1}\ll m_{\nu_2}\ll m_{\nu_3},$
this is also true for $\theta_{sol}.$ 
Therefore the prediction is expected to hold also at low scale.
However, in more general case, for example for degenerate light 
neutrinos~\cite{valle3}, 
the renormalization effects might be important.

So far we have used both the SU(5) and  SO(10) GUT relations to derive the 
prediction \rfn{prediction}. However, \rfn{prediction} can also follow from 
the SU(5) GUT constraint {\it alone}, with the additional assumption that the
phenomenological rotation by the $V_{CKM}$ is left-right symmetric (this is 
automatic for the symmetric $Y_u=Y_\nu$). Indeed, to zeroth order in $\lambda$ we have 
no left rotations and maximal right rotations diagonalizing $Y_d,$ and 
vice verse for $Y_e.$ In this basis ($Y_u,$ $Y_\nu$ are diagonal) 
we may introduce the 
corrections to order $\lambda$ as
\bea
Y_d \to  V_C Y_d V_C^\dagger,
\label{VC2}
\eea
instead of \rfn{VC}, which creates non-diagonal $V_{CKM}$ and {\it decreases}
the maximal right-rotation in $U_d$ by $\theta_C.$ Due to the SU(5) GUT relation
\rfn{su5rel}, the rotation \rfn{VC2} affects also $V_{MNS}$ and implies
 $\theta_{sol}=\pi/4 - \theta_C.$ Again, the same result is obtained as before.
This framework is simpler than the previous one since only the SU(5) 
GUT constraints are involved. However, the equality of left and right rotations
in \rfn{VC2} is an assumption replacing \rfn{so10rel}.

Extending our discussion beyond the first order in 
$\lambda$ is straightforward. 
Obviously the prediction $\theta_{atm}+\theta_{23}^{CKM}=\pi/4$
holds, just the smallness of $\theta_{23}^{CKM}\approx\lambda^2$
does not allow to test the deviation of $\theta_{atm}$ from the 
maximal. The importance of going beyond the first order in $\lambda$
is in the prediction for $\theta_{13}^{MNS}$ which should be non-zero in
order to see CP violation in the neutrino sector. Naturally we
expect (up to renormalization corrections) 
$\theta_{13}^{MNS}\sim \theta_{13}^{CKM}\sim {\cal O}(\lambda^3) $
which is, unfortunately, too small for generating observable CP violating effects
in the presently planned oscillation experiments.

We know that at least one of the simplest GUT relations, $Y_e=Y_d^T,$ must
be corrected. However, for obtaining our results we need that only the
particle mixing, which comes from the breaking on some non-Abelian flavor
symmetry, must reflect the GUT structure discussed so far. 
The masses of light quarks and leptons,
which should come from the breaking of additional U(1) flavor symmetry (otherwise
the light generations remain massless), can naturally  
differ from each other. To put it in another way, 
the flavor physics inducing \rfn{WMSSM} ``knows'' the underlying unification
structure, but after the flavor symmetry breaking this is reflected only in the 
fermion mixing and not in the eigenvalues of the Yukawa matrices.
Therefore the GUT conditions 
\rfn{su5rel} and \rfn{so10rel} are actually unnecessarily restrictive. 
For our results to be correct, 
we need that the diagonalizing matrices follow the GUT relations, 
and/or that the additional rotation by $V_{CKM}$ is left-right symmetric. 
The eigenvalues 
(quark masses) can be different (although the constraint $Y_u=Y_\nu$ is still allowed
by experimental data~\cite{BW}).  
In the following we show that this is exactly the picture what one expects to get
from a simple non-Abelian flavor model.

\mysection{Non-Abelian flavor model} 
To exemplify the ideas presented so far we need to present a model
which generates two $2\times 2$ maximal mixings from the breaking
of non-Abelian flavor symmetry, and in which fermion masses and
the mixing are not directly related to each other. Such a model
(which requires some modifications) is presented in Ref.~\cite{BHKR}.
To sketch the ideas developed in~\cite{BHKR}, let us assume that the underlying
flavor physics is based on SO(3) or SU(2) flavor symmetry.
Let us first consider two generations of fermions 
(second and third) which couple to flavons $\phi$ via
\bea
 W &=& 
(E^c\phi_E) (L\phi_{1L})  H_1 + (D^c \phi_D) (Q \phi_{1Q})  H_1 + 
(U^c\phi_U) (Q \phi_{2Q}) H_2 
\nn
&& + (N^c \phi_N) (L \phi_{2L}) H_2 +\frac{1}{2} (N^c\phi_N) M (N^c\phi_N). 
\label{Wphi}
\eea
In front of each term we implicitly assume a ${\cal O}(1/\Lambda^2)$ coefficient, where
$\Lambda$ is the flavor breaking scale.  We assume $\Lambda$ to be close to $M_{GUT}$
so that the flavon-mediated nonstandard interactions do not affect our numerical  
results. We assume that the light neutrino masses come only from the seesaw mechanism
and the flavor physics itself does not generate additional 
effective operator \rfn{effop} so that hierarchical neutrino masses can be generated
(this is not the case in~\cite{BHKR} which considered degenerate neutrino masses).
Degeneracy of light neutrinos in this context implies the U(1) breaking parameter 
of order unity.

It has been shown in~\cite{BHKR} that, with the simplest superpotentials
for the flavon fields, after symmetry breaking the flavons acquire two
types of vacuum expectation values (vevs) (writing just schematically, 
up to coefficients of order $\Lambda$)
\bea
\left(  
\begin{array}{r}
0 \\
1
\end{array}
\right)
\qquad \mrm{or} \qquad
\left(  
\begin{array}{r}
1  \\
1
\end{array}
\right).
\label{vevs}
\eea
This is a robust prediction, any deviation from this vev structure requires 
considerably more sophisticated model building.
Substituting those vevs into \Eq{Wphi}, one gets the Yukawa matrices
of the types
\bea
\left(  
\begin{array}{cc}
0 & 0 \\
0 & 1
\end{array}
\right)
\qquad \mrm{or} \qquad
\left(  
\begin{array}{cc}
0 & 1 \\
0 & 1
\end{array}
\right)
\qquad \mrm{or} \qquad
\left(  
\begin{array}{cc}
1 & 1 \\
1 & 1
\end{array}
\right).
\label{matrices}
\eea
The predictions are clear: 
$(i)$ fermion masses must be hierarchical
because one of the eigenvalues is always vanishing~\cite{singular}; 
$(ii)$ there are only vanishing or maximal flavor
mixing, depending on the corresponding flavon vev.
For example, the maximal atmospheric neutrino mixing 
and the vanishingly small  $\theta_{23}$ in the CKM matrix require
\bea
\langle \phi_{1Q}\rangle=\langle \phi_{2Q}\rangle=\langle \phi_{U}\rangle=
\langle \phi_{2L}\rangle=\langle \phi_{E}\rangle=\langle \phi_{N}\rangle=
\left(  
\begin{array}{r}
0  \\
1
\end{array}
\right) , 
\label{flavor1}
\\
\langle \phi_{1L}\rangle=\langle \phi_{D}\rangle=
\left(  
\begin{array}{r}
1  \\
1
\end{array}
\right).
\label{flavor2}
\eea
This produces particle mixing matrices $U,V$ in agreement with the GUT relations 
\rfn{su5rel}, \rfn{so10rel}. Therefore \Eq{flavor1}, \rfn{flavor2} can be considered
to be GUT constraints for the non-Abelian flavor breaking.
However, the magnitude of Yukawa couplings themselves depends
on the numerical coefficients in \rfn{Wphi}, and need not to follow the minimal
GUT relations exactly.

This is how \Eq{Wphi} generates just one $2\times 2$ 
rotation $\theta_{23}$ in each $U,V$ of \Eq{V}.
In order to generate also the maximal (1-2) mixing, one must work 
with three fermion generations and to include additional superpotential
terms for light generations into  \Eq{Wphi}. 
To give small masses to the first and second generation fermions, 
there must be additional Froggatt-Nielsen type coefficients 
weighting those terms. Details for the relevant flavon superpotentials
can be found in~\cite{BHKR}. As a result, exactly bi-maximal mixing 
with the mixing matrices consistent with the GUT relations can be produced.  
The additional rotation by the CKM matrix via \rfn{VC} or \rfn{VC2}
should occur from the mechanism beyond this model. Thus $V_{CKM}$ should 
be considered as a phenomenological parameter matrix.

Before concluding let us emphasize that the ideas presented here rely on
several untested assumptions such as small neutrino renormalization effects, 
absence of non-standard interactions, high flavour and seesaw scales etc. 
Although those requirements, in particular the one of high flavour breaking scale,
are natural in the GUT context, they can be proven wrong in new experiments
and induce important new observable effects. Degeneracy of light neutrinos in this 
scheme implies U(1) breaking parameters of order unity, and important 
renormalization effects. In those cases the results of this 
work should be reconsidered.

\mysection{Conclusions}
We argue that the (yet unknown) underlying non-Abelian flavor physics implies 
exactly bi-maximal particle mixing structure in the fermion sector, 
and that $V_{CKM}$ measures the deviation of $V_{MNS}$ from being exactly
bi-maximal. Thus, in the Wolfenstein parametrization,  $\lambda$ is the
single parameter characterizing the non-triviality of particle mixing 
both in the quark and lepton sector. We predict  
 $\theta_{sol}+\theta_{C}=\pi/4$, $\theta_{atm}+\theta_{23}^{CKM}=\pi/4$
and $\theta_{13}^{MNS}\sim \theta_{13}^{CKM}\sim {\cal O}(\lambda^3) $, 
in a good agreement with the experimental data \rfn{expprediction}.
Observable deviations from those predictions, in particular large $\theta_{13}^{MNS},$ 
allow to test the proposed scheme in the future neutrino experiments.
This prediction can follow from the SU(5) (or SU(5) and SO(10)) 
type GUT constraints for the fermion mixing matrices, and
from the structure of $V_{CKM}$ and $V_{MNS}$ in \rfn{VCKM}.
It can be considered to be 
$(i)$  a new experimental evidence for the idea of grand unification;
$(ii)$  a probe for underlying bi-maximality of the fermion mixing.
Additionally, because of (almost) vanishing $2\times 2$ sub-determinants of all 
the Yukawa matrices, this picture predicts hierarchical fermion masses 
in agreement with observations. This pattern requires both the non-Abelian
flavor symmetry breaking (which generates mixing) and the additional Abelian 
flavor symmetry breaking (which generates masses for light generations).
Based on \cite{BHKR}, we have given an example how
such a flavor structure could arise, and how it can be consistent
with the observed light quark and lepton masses (yet predicting \rfn{prediction}).

\mysection{Acknowledgment}
I would like to thank Gian Giudice and Graham Ross for discussions 
and Alessandro Strumia for providing
the results of his latest fit to neutrino oscillation data.
This work was supported by the ESF Grants 5135 and 5935, by the
EC MC contract MERG-CT-2003-503626, and by the Ministry of Education and 
Research of the Republic of Estonia.

\end{document}